\documentstyle[twoside,fleqn,espcrc2,epsf]{article}

\newcommand {\beq}{\begin{equation}}
\newcommand {\eeq}{\end{equation}}
\newcommand {\beqa}{\begin{eqnarray}}
\newcommand {\eeqa}{\end{eqnarray}}
\newcommand {\beqan}{\begin{eqnarray*}}
\newcommand {\eeqan}{\end{eqnarray*}}
\newcommand {\n}{\nonumber \\}

\newcommand {\Romannumeral}[1]{\uppercase\expandafter{\romannumeral#1}}

\newcommand {\tr} {\mbox{tr}}

\newcommand{\AmS}{{\protect\the\textfont2
  A\kern-.1667em\lower.5ex\hbox{M}\kern-.125emS}}

\title{Numerical Studies of the Double Scaling Limit \\
        in a Large $N$ Reduced Model}

\author{T. Nakajima\address{Department of Physics, University of Tokyo,
     Bunkyo-ku, Tokyo 113, Japan}%
        and 
        J. Nishimura\address{Department of Physics, Nagoya University,
     Chikusa-ku, Nagoya 464-01, Japan}
        \thanks{Speaker at the conference.}
}
       
\begin{document}

\begin{abstract}
We study the two-dimensional Eguchi-Kawai model as a toy
model of the IIB matrix model, which has been recently proposed 
as a nonperturbative definition of the type IIB superstring theory.
While the planar limit of the model is known to reproduce the
two-dimensional Yang-Mills theory,
we find through Monte Carlo simulation that the model allows 
a different large $N$ limit, which
can be considered as the double scaling limit in matrix models.
\end{abstract}

\maketitle

\section{INTRODUCTION}
Large $N$ (partially) reduced models (See Ref. \cite{Das} for a
review.) revived recently in the context of string theory
\cite{BFSS,IKKT}.
The IIB matrix model \cite{IKKT}
is a matrix model, which can be obtained by the 
dimensional reduction
of the 10D large $N$ super Yang-Mills theory to 0D (a point).
It is conjectured to provide a nonperturbative definition 
of type IIB superstring theory in the double scaling limit.
One of the interesting features of this model is that 
all the target-space coordinates come out as the eigenvalues of
the matrices.

Nonperturbative studies of the bosonic string theory 
in less than one dimension were successfully done
through the double scaling limit of the matrix model \cite{DSL}
some time ago.
On the other hand, large $N$ reduced models have been studied so far
exclusively in the planar limit,
in which the models are equivalent to the field theory before being
reduced.
Whether a large $N$ reduced model allows 
any sensible double scaling limit, which 
enables an interpretation as a string theory,
is itself a nontrivial question.

In this article, we report on our studies 
\cite{NakajimaNishimura} of 
the two-dimensional Eguchi-Kawai~(EK) model \cite{EK}, 
as a toy model of the IIB matrix model.
The model is nothing but an SU($N$) lattice
gauge theory on a $1 \times 1$ lattice
with a periodic boundary condition
and it is equivalent to an SU($N$)
lattice gauge theory on an infinite lattice
in the planar limit.
We perform a Monte Carlo simulation of the model and find
that the model indeed allows
a different large $N$ limit, which
can be considered as the double scaling limit in matrix models.

\section{EK MODEL AND PLANAR LIMIT}

The EK model is defined by the following action \cite{EK}.
\begin{equation}
  \label{red-action}
  S_{EK}=-N\beta \sum_{\mu \ne \nu=1}^{D} 
\mbox{tr} \{U_\mu U_\nu U^{\dag}_\mu U^{\dag}_\nu \}.
\end{equation}
This model has a U(1)$^{D}$ symmetry.
\begin{equation}
U_\mu \to e^{i \theta_\mu} U_\mu.
\label{usym}
\end{equation}
In Ref. \cite{EK}, it has been shown that
if the U(1)$^{D}$ symmetry is not spontaneously broken,
the model is equivalent to an SU($N$) lattice gauge theory
on an infinite lattice in the large $N$ limit,
where the coupling constant $\beta$ in the action (\ref{red-action})
is kept fixed.
This limit is referred to as the planar limit, since in this limit
Feynman diagrams with planar topology dominate.
The observable which corresponds to the
Wilson loop in the ordinary lattice gauge theory can be defined by
\begin{equation}
  \label{W-loop}
   w(C)=\frac{1}{N} \mbox{tr} \left\{
    U_\alpha U_\beta U_\gamma \cdots U_\lambda \right\},
\end{equation}
where only the direction of the links on which the link variables 
sit is maintained.

In two dimensions, 
the U(1)$^2$ symmetry is not spontaneously broken, and
therefore the EK model is equivalent to the lattice gauge theory
in the planer limit in the above sense.
The two-dimensional lattice gauge theory is solvable \cite{Gross-Witten}
in the planar limit
and indeed the exact results obtained there
have been reproduced by the EK model in the planar limit \cite{EK}.

The planar result for the expectation value of 
rectangular Wilson loops 
is given by \cite{Gross-Witten}
\begin{equation}
  \label{WilsonLoop}
  \langle W(I \times J) \rangle 
= \exp \left( -\kappa(\beta) I J \right),
\end{equation}
where
\begin{equation}
  \label{tension}
  \kappa(\beta) =\left\{
    \begin{array}{ll}
      -\log\beta & \;\;\;\mbox{for}\;\;\; \beta\le\frac{1}{2} \\
      -\log\left(1-\frac{1}{4\beta}\right) &
      \;\;\;\mbox{for}\;\;\; \beta\ge\frac{1}{2}.
    \end{array}
    \right.
\end{equation}
This result shows that the rectangular Wilson loops obey an area law
exactly for all $\beta$.
From this result, one can figure out how to fine-tune the coupling
constant $\beta$ as a function of the lattice spacing $a$ 
when one takes the continuum
limit $a \rightarrow 0$.
Since the physical area is given by $S=a^2 IJ $, we have to fine-tune
$\beta$ so that $\kappa(\beta)/a^2$ is kept fixed.
We therefore take 
\begin{equation}
\label{ptension}
a=\sqrt{\kappa(\beta)}
\end{equation}
in the following.
Since $\kappa(\beta)$ is given by eq. (\ref{tension}),
we have to send $\beta$ to infinity as we take the $a \rightarrow 0$
limit.

\begin{figure}[htbp]
  \begin{center}
    \leavevmode
    \epsfysize=6cm
    \epsffile{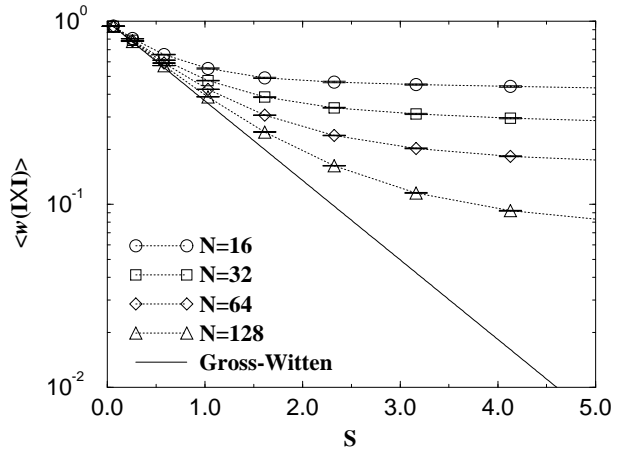}
    \caption{The expectation value of $w(I\times I)$ in the EK model 
for $\beta =4.0$ is plotted against the physical area $S=(a I)^2$.
The straight solid line represents the planar result by Gross-Witten.}
\label{nekwlp2}
  \end{center}
\end{figure}

In Fig. \ref{nekwlp2}
we plot the expectation value of square Wilson loops
in the EK model
against the physical area $S=(aI)^2$
for $\beta=4.0$ with $N=16,32,64$ and $128$.
One can see that the data points approach the planar result
from above monotonically as we increase $N$.
This is not the case when the boundary condition is
twisted \cite{NakajimaNishimura}.

\section{DOUBLE SCALING LIMIT}
We consider the EK model as a toy model of the IIB matrix model.
Just as in Ref. \cite{IKKT}, we make the T-duality transformation,
when we interpret the EK model as a string theory.
Since the EK model can be considered to be
defined on a unit cell of size $a$ 
with a periodic boundary condition,
it can be considered, after the T-duality transformation,
as a string theory in the two-dimensional space time
compactified on a torus of size $1/a$.

The observables we consider in this section are the Wilson loops
\beq
\mbox{tr} \left\{ U_\alpha U_\beta U_\gamma \cdots U_\lambda \right\},
\label{stringf}
\eeq
where the suffix $\alpha,\beta,\cdots,\lambda$ runs over
$\pm 1$,$\pm 2$ and $U_{-\alpha}$ ($\alpha > 0$) is defined by
$U_{-\alpha} = U_{\alpha}^{\dagger}$.
The winding number of the Wilson loop is given by
$n_\mu=\hat{\alpha}_\mu+\hat{\beta}_\mu+\cdots + \hat{\lambda}_\mu$,
where $\hat{\alpha}_\mu$ denotes a unit vector in the $\alpha$ direction.
Note that the observables (\ref{W-loop})
considered in the previous section can be viewed as
the Wilson loops in the EK model with $n_\mu =0$,
while in this section we consider those with
$n_\mu \ne 0 $ as well.
After the T-duality transformation, 
the winding of the Wilson loops represents
the momentum distribution of the string in the two-dimensional space time.

In what follows, we consider 
\beqa
O^{(n)}_{\alpha\beta}(L) &=& \tr(U_\alpha ^L U_\beta ^L 
                U_{-\alpha} ^L U_{-\beta} ^L  ) \n
O^{(w)}_{\alpha}(L) &=& \tr(U_\alpha ^L),
\label{typicalO}
\eeqa
which give typical Wilson loops in the non-winding sector
and the winding sector respectively.
When we take the $a\rightarrow 0$ limit,
we have to send $L$ to the infinity by
fixing $aL$.

\begin{figure}[htbp]
  \begin{center}
    \leavevmode
    \epsfysize=6cm
    \epsffile{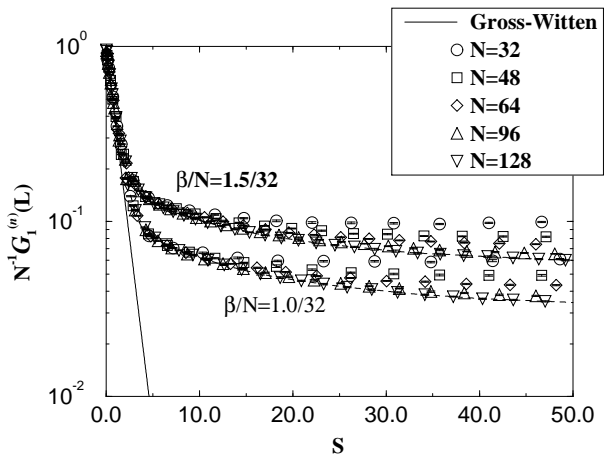}
    \caption{
The one-point function of the Wilson loops
in the non-winding sector $N^{-1}G_1^{(n)}(L)$ 
in the EK model is plotted
against $S=(aL)^2$ for 
the sets of $\beta$ and $N$ with $\beta/N$ fixed to 
$1.0/32$ and $1.5/32$.
The straight solid line represents the planar result by Gross-Witten.
The dashed lines, which connect the data for $N=128$,
are drawn to guide the eye.}
\label{dblwlp}
  \end{center}
\end{figure}

Let us examine the one-point function
of Wilson loops in the non-winding sector, namely
$G_1^{(n)}(L) = \langle O^{(n)}_{12}(L) \rangle $.
The observable $\langle w(L \times L) \rangle $ 
considered in the previous section
is nothing but $N^{-1} G_1^{(n)}(L)$.
We have seen in Fig. \ref{nekwlp2}
how the data for $N^{-1}  G_1^{(n)}(L)$ 
approach the planar limit when we increase $N$ with fixed $\beta$.
We find that 
$N^{-1}  G_1^{(n)}(L)$ scales in the large $N$ limit
by fixing $\beta/N$.
In Fig. \ref{dblwlp} we show the data
for $\beta/N =1.0/32$ and $\beta/N =1.5/32$.
$N$ is taken to be $32,48,64,96$ and $128$.

From this observation, it is natural to expect that
the double scaling limit of the EK model can be taken by
fixing $\beta/N$, which corresponds to 
the string coupling constant $g_{str}$.
The planar limit corresponds to $g_{str}=0$.
We have also observed 
\cite{NakajimaNishimura}
a clear scaling behavior 
for other Green functions; {\it e.g.},
\begin{equation}
  \label{opn2pt}
  G_2^{(w)}(L) = 
\langle O_\mu^{(w)}(L) O_{-\mu}^{(w)}(L) \rangle . 
\end{equation}

\section{CONCLUSION AND DISCUSSION}

To summarize, 
our numerical results strongly suggest the existence
of a double scaling limit in the 2D EK model.
In this limit, 
the density of eigenvalues is constant on a physical scale.
As we mentioned above, 
the extent of space time is given by $R=1/a$.
Since we have $N$ eigenvalues distributed in a two-dimensional
space time with this extent, the average density is given by
$N/R^2 = N a^2 $, which is constant in the double scaling limit.
This fact is natural from a string theoretical point of view,
since it means that there are only finite dynamical degrees 
of freedom on average in a finite region of space time.

The fact that a sensible double scaling limit can be taken for a
large $N$ reduced model is itself
encouraging for the study of a nonperturbative formulation of 
superstring theory through the IIB matrix model.
We hope to report on numerical studies of the IIB matrix model
in future publications.



\begin{thebibliography}{9}
\bibitem{Das} S.R. Das, 
Rev. Mod. Phys. {\bf 59} (1987) 235.
\bibitem{BFSS} T. Banks, W. Fischler, S.H. Shenker and L.Susskind,
Phys. Rev. {\bf D55} (1997) 5112.
\bibitem{IKKT}
N. Ishibashi, H. Kawai, Y. Kitazawa and A. Tsuchiya,
Nucl. Phys. {\bf B498} (1997) 467.
\bibitem{DSL}
E. Br\'ezin and V.A. Kazakov, 
Phys. Lett. {\bf B236} (1990) 144. \\
M.R. Douglas and S.H. Shenker,
Nucl. Phys. {\bf B335} (1990) 635. \\
D.J. Gross and A.A. Migdal, 
Phys. Rev. Lett. {\bf 64} (1990) 127.
\bibitem{NakajimaNishimura}
T. Nakajima and J. Nishimura, Nucl. Phys. {\bf B528} (1998) 355.
\bibitem{EK} 
T. Eguchi and H. Kawai, 
Phys. Rev. Lett. {\bf 48} (1982) 1063. 
\bibitem{Gross-Witten} D.J. Gross and E. Witten,
Phys. Rev. {\bf D21} (1980) 446.

\end{thebibliography}
\end{document}